\documentclass[12pt, letterpaper]{article}
\usepackage{geometry}
\usepackage{amsmath}
\usepackage{graphicx}
\usepackage{float}
\usepackage{subcaption}
\usepackage{amssymb}
\newgeometry{vmargin={18mm}, hmargin={18mm,18mm}}
\providecommand{\keywords}[1]
{
  \small	
  \textbf{\textit{Keywords---}} #1
} 

\begin{document}
\title{Interaction and adiabatic evolution of orthodromic and antidromic impulses in the axoplasmic fluid}

\author{O. Pav\'on-Torres$^{a}\footnote{E-mail: omar.pavon@cinvestav.mx (corresponding author)}$, M. A. Agüero-Granados$^{b}$, M. E. Maguiña-Palma$^{c}$}
\date{%
\small{$^{a}$ Physics Department, Cinvestav, POB 14-740, 07000 Mexico City, Mexico\\ %
$^{b}$ Facultad de Ciencias, Universidad Aut\'onoma del Estado de M\'exico, \\ Instituto literario 100, Toluca 5000, M\'exico.\\
$^{c}$ Universidad Norbert Wiener, Jr. Larrabure y Unanue 110, Av. Arequipa 440,\\ Lima, Lima 01.
}}
\maketitle


\begin{abstract}
Unlike expected from the Hodgkin-Huxley model predictions, in which there is annihilation once orthodromic and antidromic impulses collide, the Heimburg-Jackson model demonstrates that both impulses penetrate each other as it has been shown experimentally. These impulses can be depicted as low amplitude nonlinear excitations in a weakly dissipative soliton model described by the damped NLSE. In view of the above, the Karpman-Solov'ev-Maslov perturbation theory turns out to be ideal to study the interaction and adiabatic evolution of orthodromic and antidromic impulses once axoplasmic fluid is present.
\\
\\
\keywords{Orthodromic-antidromic impulses; Solitons collision; Perturbation theory.}
\end{abstract}



\section{Introduction}

Collision events of impulses within neurons are fundamental for comprehending neural activity and the transmission of signals \cite{rr1, rr2, rr3, jojo}. Historically, these collision events dates back to 1949 when Tasaki conducted a pioneering series of experiments \cite{tasa}. These experiments conclusively established that when two impulses collide, they undergo mutual annihilation, corroborating theoretical predictions from the Hodgkin-Huxley (HH) model \cite{r00, r01}. Nevertheless, a recent collision experiments using the ventral cords of earthworm \textit{Lumbricus terrestris} and the abdominal ventral cord of lobster \textit{Homarus americanus} have yielded intriguing results. These experiments demonstrated that when two impulses generated simultaneously in opposite directions (orthodromic and antidromic) collide head-on, they do not undergo mutual annihilation, as previously expected \cite{r12}. Instead, they penetrate each other and emerge from the collision unaltered in shape and velocity. The latter experiment together with local compression of the biomembrane resulting in liquid-to-gel state transition in nerve has joined to a long track record of phenomena which cannot be described by the HH model.  In this sense, the thermodynamic soliton theory of the nervous impulse proposed by Heimburg and Jackson \cite{heim1, heim2, heim3} has once again captured the attention \cite{r0, r5, r8, o1}. Standing aside from the Hodgkin-Huxley perspective, the thermodynamic soliton theory suggests that colliding pulses do not result in annihilation but rather facilitate near-lossless penetration. The model suggests that the dominant effect in nerve signalling might be mechanical rather than electrical and the Action Potential (AP) is an adiabatic phenomenon. Heating can destroy the lateral order of molecules, which absorbs heat and leads to a swelling of the structure leading to changes in volume and area compressibility of the membrane \cite{o2}. As a result, in the proximity of the transition, the speed of sound becomes contingent upon the mass per unit area, while the AP is coupled with a longitudinal pressure wave within the cell membrane. This pressure wave is a local travelling liquid-to-gel phase transition of the lipid membrane that can propagate long distances without significant energy dissipation. In particular, two forms of this wave within the nerve have been identified: bright soliton, for gel state, and dark soliton, for the liquid state \cite{r9, r17}. Both profiles of solitons are obtained by considering the weak formulation of the Heimburg-Jackson model, a similar approach has been considered to reduce a modified diffusion equation to a damped nonlinear Schrödinger equation \cite{r1}. In tandem with these studies, a variety of soliton forms have recently been obtained using the Hirota method \cite{multi} and the $Exp(-\varphi(\xi))$-expasion method \cite{r7}. The thermodynamic soliton theory of nervous impulses provides a robust framework for understanding the formation of soliton structures in biomembranes and their deep implications in biophysics and medical applications. Moreover, it serves as a comprehensive platform for investigating penetration phenomena and offers invaluable insights into the influence of axoplasmic fluid on both orthodromic and antidromic impulses.


Thus, in the present study, our aim is to offer a robust explanation for the behavior of both orthodromic and antidromic impulses in the axoplasmic fluid. To achieve this, we employ the thermodynamic soliton theory in conjunction with the Karpman Maslov-Solov'ev soliton perturbation theory \cite{r15, r16, r20}.  These theoretical frameworks enable us to delve deeply into the dynamics of neural impulses, shedding light on their intricate behavior and interactions within the nervous system. We contrast our theoretical predictions with experimental data and numerical simulations based on a variational method \cite{r11}. The present work is divided in the following sections. In section 2, a brief overview and the general conceptions of the Heimburg-Jackson model will be presented. Also, we outlined the general procedure to reduce the Heimburg-Jackson dissipative equation to a perturbed Nonlinear Schrödinger Equation (pNLSE) by considering nerve impulses as low amplitude nonlinear excitations in a weakly dissipative soliton model. The section 3 is devoted to present and implement the basic ideas of the Karpman-Maslov-Solov’ev soliton perturbation theory to study orthodromic and antidromic impulses collision. Subsequently, in the following section the effect of mechanical changes in the nerve, e.g. swelling or stretching of the nerve or the viscous elastic axoplasmic fluid will be considered. In addition, a discussion and comparison of the different approaches will be performed. Finally, in the last section a perspective analysis and the conclusion of our work will be presented.

\section{Thermodynamic soliton theory of the nervous impulse}

\subsection{The Heimburg-Jackson model}

Electrical signals, encompassing both action potentials and ion currents, serve as the cornerstone for information transmission within the nervous system. The HH model \cite{r00, r01} and the FitzHugh-Nagumo (FHN) model \cite{nag1, nag2} have been fundamental in modelling these electrical phenomena, offering distinct yet complementary perspectives on neural dynamics. The FHN model, while proficient in simulating the primary characteristics of the AP, is limited to a single ion current. When the intricate effects of individual ion currents become essential, the Hodgkin-Huxley (HH) model or its adapted versions prove indispensable. For a comprehensive representation of dynamic phenomena in nerve fibres, we start our analysis with the wave


\begin{equation}
\dfrac{\partial^{2}U}{\partial t^{2}}=\dfrac{\partial}{\partial x}\left(c^ {2}\dfrac{\partial U}{\partial x}\right)\label{panda00},
\end{equation}
which is a balance of momentum where $U=\rho^ {A}-\rho_{0}^ {A}$ is the density change between the gel-state density $(\rho^{A})$ and the fluid-state density $(\rho^{A}_{0})$. 
The first assumption of the Heimburg-Jackson model establishes a fundamental connection between the velocity $c$ and the compressibility inherent in the lipid composition of the circular biomembrane. It is assumed that
\begin{equation}
c^{2}=c_{0}^ {2}+pU+qU^ {2},\label{panda0}
\end{equation}
where $c_{0}$ is the speed of sound during the phase transition, $p$ and $q$ are nonlinear elastic properties of the membrane determined from the experiments as follows
\begin{equation}
c_{0}^ {2}=\dfrac{1}{K_{s}^ {A}\rho_{0}^ {A}}; \quad p= -\dfrac{1}{\left(K_{s}^ {A}\rho_{0}^ {A}\right)^ {2}}; \quad q= \dfrac{1}{\left(K_{s}^ {A}\rho_{0}^ {A}\right)^ {3}},
\end{equation}
being $K_{s}^ {A}$ the lateral compressibility of the axon. Replacing Eq. (\ref{panda0}) in Eq. (\ref{panda00}) the Heimburg-Jackson model turns out to be
\begin{equation}
\dfrac{\partial^{2}U}{\partial t^{2}}=\dfrac{\partial}{\partial x}\left[\left(c_{0}^ {2}+pU+qU^ {2}\right)\dfrac{\partial U}{\partial x}\right]-h\dfrac{\partial ^ {4}U}{\partial x^ {4}}, \label{panda}
\end{equation}
with the added higher order $-h\partial^{4} U/\partial x^{4}$ term being responsible for dispersion. Finally, incorporating the effect of friction as a mixed derivative in Equation (\ref{panda}) leads to
\begin{equation}
\dfrac{\partial^{2}U}{\partial t^{2}}=\dfrac{\partial}{\partial x}\left[\left(c_{0}^ {2}+pU+qU^ {2}\right)\dfrac{\partial U}{\partial x}\right]+\alpha\dfrac{\partial ^ {2}}{\partial x^ {2}}\left(\dfrac{\partial U}{\partial t}\right)-h\dfrac{\partial ^ {4}U}{\partial x^ {4}}, \label{panda1}
\end{equation}
the incorporation of the additional term is possible due to the similarity between the nerve fibre and the rods. A similar term is responsible for inertial effects in rods and it can be related to inertia of the lipids in biomembranes \cite{r0}. 

\subsection{The damped nonlinear Schrödinger equation}
Before continue, let us express Eq. (\ref{panda1}) in terms of dimensionless variables as follows

\begin{equation}
\dfrac{\partial^{2}u}{\partial \tau^{2}}=\dfrac{\partial}{\partial z}\left[\left(1+\tilde{p}U+\tilde{q}U^ {2}\right)\dfrac{\partial u}{\partial z}\right]+\mu\dfrac{\partial ^ {2}}{\partial z^ {2}}\left(\dfrac{\partial u}{\partial \tau}\right)-\dfrac{\partial ^ {4}U}{\partial z^ {4}}, \label{panda2}
\end{equation}
with
\begin{equation}
u=\dfrac{U}{\rho_{0}^{A}}, \quad z=\dfrac{c_{0}x}{\sqrt{h}}, \quad \tau=\dfrac{c_{0}^{2}t}{\sqrt{h}}
\end{equation}
and the new parameters being 
$$\mu=\dfrac{\alpha}{\sqrt{h}}, \quad \tilde{p}=\dfrac{\rho_{0}^{A}}{c_{0}^{2}}p \quad \tilde{q}=\dfrac{\left(\rho_{0}^{A}\right)^ {2}}{c_{0}^{2}}q$$

As we have mentioned above, orthodromic and antidromic impulses can be modelled as low amplitude nonlinear excitations in a weakly dissipative soliton model. Thus, we implement a procedure to reduce the Heimburg-Jackson diffusion equation to a damped nonlinear Schrödinger equation (see \cite{r1} and references therein). We introduce a small parameter $\epsilon\ll 1$, and substitute $\tilde{p} \to \epsilon^{2}\tilde{p}$, $\tilde{q} \to \epsilon^{2}\tilde{q}$, $\mu \to \epsilon \mu$ in Eq. (\ref{panda2}).  Followed by a change of variables $y=\epsilon (z-\tau)$ and $s=\epsilon^{3}\tau$ we obtain the Burges-Korteweg-de Vries (BKdV) equation
\begin{equation}
\dfrac{\partial u}{\partial s}+\dfrac{1}{2}(a\tilde{p}u+a^{2}\tilde{q}u^{2})u_{y}-\dfrac{1}{2}\dfrac{\partial^ {3}u}{\partial y^{3}}=a^{2}\dfrac{\mu}{2}\dfrac{\partial^ {2}u}{\partial y^{2}}.
\end{equation}
We use a multiple-scale expansion method \cite{mirage1, r10}. The method involves introducing two time and spatial scales, i.e., the fast time and spatial scales for oscillations and the slow time and spatial scales for the envelope amplitude. Thus, we introduce new time scale $s_{i}=a^ {i}s$ and space scale $y_{i}=a^ {i}y$, and assume a solution in terms of the Poincar\'e type expansion
\begin{equation}
u(y, s)=\sum_{i=0}^ {\infty}a^{i}u(s_{0}, s_{1}, s_{2}, y_{0}, y_{1}),
\end{equation}
in which $s_{i}$ and $y_{i}$ is treated as an independent variable. This leads to a perturbation series of operators for all independent variables 
\begin{equation}
\dfrac{\partial}{\partial s}=\dfrac{\partial}{\partial s_{0}}+a\dfrac{\partial}{\partial s_{1}}+a^ {2}\dfrac{\partial}{\partial s_{2}}+\dots \label{wom1}
\end{equation}
The solution of the original problem will only be obtained, if a multidimensional space is generated by the new sets of variables $s_{i}$ and $y_{i}$ with the explicit form given by
\begin{equation}
s_{0}=s, \quad s_{1}=as, \quad s_{2}=a^ {2}s \label{ladworm}.
\end{equation}
Replacing the above operator (\ref{wom1}) and their counterpart for $y$ into the different terms of the BKdV equation and group the terms of the same order of $a$ to obtain a system of equations. Each of these equations will correspond to each approximation having harmonics of a specific order. Picking the operators:

\begin{equation}
\dfrac{\partial^ {2}}{\partial y^{2}}=\dfrac{\partial^{2}}{\partial y_{0}^{2}}+2a\dfrac{\partial^ {2}}{\partial y_{0}\partial y_{1}}+a^{2}\dfrac{\partial^ {2}}{\partial y_{1}^{2}},\label{wom3}
\end{equation}
and writing $u$ as a perturbative series, and consider only terms to the first order of $a$, i.e., 
\begin{equation}
u=Ae^{i\theta}+A^{*}e^{-i\theta}+a(C+Be^{2i\theta}+B^{*}e^{-2i\theta}), \label{wom4}
\end{equation}
where $\theta=(ky_{0}-\omega s_{0})$ and the amplitudes $A$, $B$ and $C$ correspond to $(s_{1}, y_{1}, y_{2})$. Substituting Eqs. (\ref{wom1})-(\ref{wom4}) in Eq. (\ref{panda2}), and looking for relations between terms of same orders in $a$ with terms in $e^{\pm i \theta}$, $e^{\pm 2i \theta}$ without an exponential dependence set to zero. To the order $a^{0}$ the annihilation of terms in $e^ {\pm i \theta}$ gives the dispersion relation of linear waves, i.e., 
\begin{equation}
\omega=\dfrac{k^{3}}{2}.
\end{equation}
To the order $a^{1}$, cancellation of terms in $e^{\pm i \theta}$ gives 
\begin{equation}
\dfrac{\partial A}{\partial s_{1}}+v_{g}\dfrac{\partial A}{\partial y_{1}}=0 \label{lw1},
\end{equation}
where $v_{g}$ is the group velocity defined as
\begin{equation}
v_{g}=\dfrac{\partial \omega}{\partial k}=\dfrac{3k^{2}}{2}.
\end{equation}
Setting terms in $e^{\pm 2i \theta}$ to zero we have
\begin{equation}
B=-\dfrac{pk}{6k^{3}}A^{2}.
\end{equation}
To the second order, the terms with zero exponential dependence yield
\begin{equation}
\dfrac{\partial C}{\partial s_{1}}-\dfrac{p}{2}\dfrac{\partial |A|^ {2}}{\partial y_{1}}=0.
\end{equation}
Now, considering the transformations $y=(y_{1}-v_{g}s_{1})$ and $\tau=s_{1}$, and comparing the result with Eq. (\ref{lw1})  we obtain
\begin{equation}
C=\dfrac{p}{3k^ {2}}|A|^{2}
\end{equation}
To the second order perturbation,  the terms of order $e^{\pm i \theta}$ finally give the damped nonlinear Schrödinger equation we were looking for
\begin{equation}
i\dfrac{\partial A}{\partial s_{2}}+P\dfrac{\partial^{2} A}{\partial y_{1}^{2}}+Q|A|^{2}A+iRA=0 \label{schr1},
\end{equation}
which describes the evolution of the envelope amplitude of the density pulse $u$ in the one-dimensional cylindrical nerve axon. Here the nonlinearity or (self-trapping), the damping and the dispersive coefficients $Q$, $R$ and $P$, respectively, are real and defined in terms of membrane parameters as

\begin{equation}
Q=\left(\dfrac{\rho_{0}^{A}}{2c_{0}}\right)\left(\dfrac{p^{2}}{3c_{0}^{2}k^{2}}-k{q}\right), \quad P=\dfrac{3k}{2}, \quad R=\dfrac{\alpha k^ {2}}{2\sqrt{h}}\label{agon0}.
\end{equation}
The inclusion of the imaginary term in the nonlinear Schrödinger equation is keynote in depicting the damping of amplitude, illustrating the irreversible nature of time evolution. However, it doesn't directly translate in energy dissipation. This damping term is often attributed to the elastic viscous properties of the medium under consideration, with the axoplasmic fluid serving as a prime example in this context.

\medskip

We must outline that, under the present approach, the study of the pressure wave inside the cell membrane and its interaction with the axoplasmic fluid can be done straightforward. From the unperturbed nonlinear Schrödinger equation, i.e, for the case when the last term of Eq. (\ref{schr1}) is equal to zero we have the classic soliton structures. As was mentioned above these mentioned nonlinear structures, bright and dark solitons, arise from the phase transitions of the lipid membrane \cite{r9, r17}. In addition, the present reduction allows to consider orthodromic and antidromic impulses as solitons in the Heimburg-Jackson model. The latter opens up the possibility to study them in the scheme of Karpman-Maslov-Solov'ev perturbation theory and to give an insight about their behaviour. As it is well known from soliton definition once they collide frontally they penetrate each other with no significant change. Thus, the present model presents an explanation of experimental observations recently obtained \cite{r12}.

\section{Adiabatic evolution of orthodromic and antidromic impulses}

Physiologically, the beginning of an AP typically occurs at the axon hillock within neurons. However, intriguingly, in both vertebrate and invertebrate nerve cells, the generation of APs isn't confined solely to this region. Remote sites from the axon hillock can also trigger APs. Pulse propagation follows two principal pathways: orthodromic, where impulses travel from the soma towards synaptic terminals, and antidromic, where propagation moves in the opposite direction. Both orthodromic and antidromic impulse propagation can be elicited through electrical stimulation, either near the soma or distal parts of the axon, respectively. Notably, the simultaneous occurrence of orthodromic and antidromic pulses can result in collision events, underscoring the intricacies of neural signal transmission.
In order to study the interaction conditions and adiabatic evolution of these impulses we use the generalization of the Karpman-Maslov-Solov'ev soliton perturbation theory \cite{r15, r16, r20}. To start our analysis, let us rewrite the perturbed nonlinear Schrödinger equation with normalized parameters. Henceforth, fixing $P=1/2$ and $Q=1$ in Eq. (\ref{agon0}) and going back to our first convention, that is, $s_{2}\to t$ and $y_{1}\to x$ in Eq. (\ref{schr1}). Consequently, we have the following simplified version our perturbed equation
\begin{equation}
iA_{t}+\dfrac{1}{2}A_{xx}+|A|^{2}A=i\epsilon (A), \label{misha1}
\end{equation}
with $\epsilon (A)$ being a small parameter ($\epsilon\ll 1$). We look for a solution of the form
\begin{equation}
A(x, t)=\eta(t)sech[\eta(t)(x-q(t))]exp[i\phi(t)-i\delta(t)x], \label{m1}
\end{equation}
where $\eta(t)$, $q(t)$, $\phi(t)$ and $\delta(t)$ are unfixed parameters corresponding to amplitude, position of the mass center, phase and velocity of soliton. These physical parameters are slowly varying in time leading to modification of the shape and the velocity of the soliton. The original idea of Karpman and Solov'ev considered the interaction between solitons as a slow deformation of the basic parameters of the soliton (amplitude, position, phase and velocity). Based on inverse scattering they found out that the temporal dependence of the parameters of Eq. (\ref{m1})

\begin{subequations}
\begin{equation}
\dfrac{d\eta(t)}{dt}=\dfrac{1}{\eta(t)}Re\int_{-\infty}^{\infty}\epsilon (A) A^ {*}(z)dz, \label{su1}
\end{equation}
\begin{equation}
\dfrac{d\delta(t)}{dt}=\dfrac{1}{\eta(t)}Im\int_{-\infty}^{\infty}\epsilon (u)tanhz A^{*}(z)dz, \label{su2}
\end{equation}
\begin{equation}
\dfrac{dq(t)}{dt}=-\delta(t)+\dfrac{1}{\eta^{3}(t)}Re\int_{-\infty}^{\infty}\epsilon (A)z A^{*}(z)dz, \label{su3}
\end{equation}
\begin{equation}
\dfrac{dq(t)}{dt}=\dfrac{1}{\eta^{2}(t)}Im\int_{-\infty}^{\infty}\epsilon (A)[1-ztanhz] u*(z)dz+\dfrac{1}{2}(\eta^ {2}(t)-\delta^{2}(t))+q\dfrac{d\delta(t)}{dt}, \label{su4}
\end{equation}
\end{subequations}
with $z=\eta [x-q(t)]$. To study the interaction of orthodromic and antidromic impulses (solitons) we look for a solution of equation (\ref{misha1}), thus, we describe these impulses as a linear superposition of two solitons which represent the left and right wave propagation, respectively, 

\begin{equation}
A(x, t)=A_{1}(x, t)+A_{2}(x, t).
\end{equation}

In general, the varying thickness of the axon along its length results in distinct shapes for orthodromic and antidromic impulses, leading to differences in their waveforms. Both impulses can adopt several soliton-like profiles, particularly, in this work we restrict ourselves to bright-shape soliton profile impulses. The previous consideration wasn't merely for simplicity's sake; it has demonstrated a rapid convergence to an analytical stationary bright soliton solution as the number of Gaussians increases, particularly evident when two solitons collide. \cite{r11}. Consequently, every single soliton solution can be expressed by 

\begin{equation}
A_{1, 2}(x, t)=\eta_{1, 2}(t)sech[\eta_{1, 2}(t)(x-q_{1, 2}(t))]exp[i\phi_{1, 2}(t)-i\delta_{1, 2}(t)x].\label{misha2}
\end{equation}
Once we replace Eq. (\ref{misha2}) in Eq. (\ref{misha1}) we have a system of coupled perturbed nonlinear Schrödinger equations
\begin{subequations}
\begin{equation}
i\dfrac{\partial A_{1}}{\partial t}+\dfrac{1}{2}\dfrac{\partial^{2} A_{1}}{\partial x^{2}}+|A_{1}|^{2}A_{1}=i\epsilon (A_{1})-2|A_{1}|^ {2}A_{2}-A_{1}^{2}A_{2}^{*}\label{misha3}
\end{equation}
\begin{equation}
i\dfrac{\partial A_{2}}{\partial t}+\dfrac{1}{2}\dfrac{\partial^{2} A_{2}}{\partial x^{2}}+|A_{2}|^{2}A_{2}=i\epsilon (A_{2})-2|A_{2}|^ {2}A_{1}-A_{2}^{2}A_{1}^{*}\label{misha4}
\end{equation}
\end{subequations}
the right side terms of Eqs. (\ref{misha3}) and (\ref{misha4}) describe the nonlinear interaction of the orthodromic and antidromic impulses. Despite we focus on these specific soliton-profile solutions, the number of studies devoted to exploring interactions among solitons in coupled nonlinear Schrödinger equations has significantly grown in recent years \cite{mer1, mer2, mer3, r14}. 

\subsection{Conditions to orthodromic-antidromic impulses interaction}

Let us introduce the following parameters which physically characterize to the orthodromic and antidromic impulses: the mean amplitude $\eta$; the amplitude differences $\Delta \eta$; the mean velocity $\delta$; the velocities difference $\Delta\delta$; separation between solitons $q$ and the phase difference $\phi$. 

\begin{equation*}
\eta=\dfrac{1}{2}(\eta_{1}+\eta_{2}); \quad \Delta \eta=\eta_{2}-\eta_{1};
\end{equation*}
\begin{equation*}
 \delta=\dfrac{1}{2}(\delta_{1}+\delta_{2}); \quad \Delta\delta=\delta_{2}-\delta_{1}
\end{equation*}
\begin{equation*}
q=\dfrac{1}{2}(q_{2}-q_{1}); \quad \phi=\phi_{2}-\phi_{1}.
\end{equation*}

Furthermore, we take into account that the separation between the impulses is notably significant ($q_{2}\gg q_{1}$) and their velocities and amplitudes are essentially identical. That means

\begin{equation}
|\Delta \eta q| \ll 1 \ll\eta q, \quad |\phi q|\ll 1, 
\end{equation}

Thus, the complete perturbation of (\ref{misha3}) and (\ref{misha4}) considering the damping term $i\epsilon(A)=i\gamma A$ takes the form

\begin{equation}
\epsilon(A_{m})A_{m}^{*}=(\gamma+2iA_{m}^{*}A_{n}+iA_{m}A_{n}^{*})|A_{m}|^ {2} \label{su5}
\end{equation}
with $m, n=1, 2$ with $m\neq n$.
Therefore, replacing Eq.(\ref{su5}) in the Karpman-Maslov system of equations (\ref{su1}-\ref{su4})

\begin{subequations}
\begin{equation}
\dfrac{d\eta_{k}(t)}{dt}=2\gamma\eta(t)+(-1)^{k}4\eta^ {3}(t)exp(-2\eta(t)q(t))\sin\phi \label{su6}
\end{equation}
\begin{equation}
\dfrac{d\delta_{k}(t)}{dt}=(-1)^{k}4\eta^ {3}(t)exp(-2\eta(t)q(t))\cos\phi\label{su7}
\end{equation}
\begin{equation}
\dfrac{dq_{k}(t)}{dt}=-\delta_{k}(t)-2\eta(t)exp(-2\eta(t)q(t))\sin\phi\label{su8}
\end{equation}
\begin{equation}
\dfrac{d\phi_{k}(t)}{dt}=6\eta^ {2}(t)exp(-2\eta(t)q(t))\cos\phi+2\eta (t)\delta_{k}\eta(t)exp(-2\eta(t)q(t))\sin\phi+\dfrac{1}{2}(\eta_{k}^{2}(t)+\delta_{k}^{2}(t))\label{su9}
\end{equation}
\end{subequations}

Eqs. (\ref{su6}-\ref{su9}) enable us to generalize the Karpman-Solov'ev system of equations for the differences of the impulse parameters defined by the impulse parameter differences

\begin{subequations}
\begin{equation}
\dfrac{\Delta \eta(t)}{dt}=8\eta^{3}(t)exp(-2\eta(t)q(t))\sin\phi+2\gamma \Delta \eta (t), \label{sus1}
\end{equation}
\begin{equation}
\dfrac{d\Delta \delta}{dt}=8\eta^{3}(t)\exp(-2\eta(t)q(t))\cos\phi,\label{sus2}
\end{equation}
\begin{equation}
\dfrac{dq(t)}{dt}=-\dfrac{1}{2}\Delta \delta,\label{sus3}
\end{equation}
\begin{equation}
\dfrac{d\phi}{dt}=\eta(t)\Delta \eta(t).\label{sus4}
\end{equation}
\end{subequations}
Eqs. (\ref{sus1})-(\ref{sus4}) follow the equations

\begin{subequations}
\begin{equation}
\dfrac{d^{2}q(t)}{dt^{2}}=-4\eta^ {3}(t)exp(-2\eta(t)q(t))\cos\phi
\end{equation}
\begin{equation}
\dfrac{d^{2}\phi(t)}{dt^{2}}=8\eta^ {4}(t)exp(-2\eta(t)q(t))\sin\phi+4\gamma \eta \Delta \eta(t).
\end{equation}
\end{subequations}

Introducing the complex function $Y=\Delta \delta(t)+i\Delta \eta (t)$, combining Eqs. (\ref{sus1})-(\ref{sus4}) and differentiating respect to time we obtain 

\begin{equation}
\dfrac{d^ {2}Y}{dt^ {2}}=8\eta^{3}(t)exp(-2\eta(t)q(t)+i\phi)\left(-2q(t)\dfrac{d\eta}{dt}+\eta Y\right)+24exp(-2\eta q+i\phi)\eta^{2}\dfrac{d\eta}{dt}+2i\gamma\dfrac{d\Delta\eta}{dt}.
\end{equation}
Assuming $\gamma=\gamma_{0}$ in which $\gamma_{0}\ll 1$ we come up to the simplified equation 
\begin{equation}
\dfrac{d^{2}Y}{dt^ {2}}=\dfrac{1}{2}\eta(t)\dfrac{d(Y^{2})}{dt} \label{c1}
\end{equation}

For the simplest case of orthodromic and antidromic impulses interaction with no additional axoplasmic fluid ($\gamma(t)=0$) we have that $\eta(t)=\eta_{0}$ and this allows the integration of Eq. (\ref{c1}) leading to 
\begin{equation}
Y=-C\tanh\left(\dfrac{1}{2}C\eta_{0}t-\alpha\right)
\end{equation}
where $\alpha$ and $C$ are a complex constants. We propose $C$ as

\begin{equation}
C^{2}=Y^{2}-16\eta_{0}^{2}exp(-2\eta_{0}q+i\phi) \label{george1}
\end{equation}
Thus, the differences of the velocities and amplitudes of the solitons can be obtained 
\begin{equation}
\Delta \delta=\dfrac{-C_{1}\sinh(C_{1}\eta_{0}t-2\alpha_{1})+C_{2}\sin(C_{2}\eta_{0}t-2\alpha_{2})}{\cosh(C_{1}\eta_{0}t-2\alpha_{1})+\cos(C_{2}\eta_{0}t-2\alpha_{2})}
\end{equation}
\begin{equation}
\Delta \eta=\dfrac{-C_{2}\sinh(C_{1}\eta_{0}t-2\alpha_{1})-C_{1}\sin(C_{2}\eta_{0}t-2\alpha_{2})}{\cosh(C_{1}\eta_{0}t-2\alpha_{1})+\cos(C_{2}\eta_{0}t-2\alpha_{2})}
\end{equation}
and (\ref{sus3}) and (\ref{sus4}) can be rewritten as
\begin{equation}
\dfrac{dq}{dt}=\dfrac{1}{2}\dfrac{C_{1}\sinh(C_{1}\eta_{0}t-2\alpha_{1})-C_{2}\sin(C_{2}\eta_{0}t-2\alpha_{2})}{\cosh(C_{1}\eta_{0}t-2\alpha_{1})+\cos(C_{2}\eta_{0}t-2\alpha_{2})}\label{d1}
\end{equation}
\begin{equation}
\dfrac{d\phi}{dt}=\eta\dfrac{-C_{2}\sinh(C_{1}\eta_{0}t-2\alpha_{1})-C_{1}\sin(C_{2}\eta_{0}t-2\alpha_{2})}{\cosh(C_{1}\eta_{0}t-2\alpha_{1})+\cos(C_{2}\eta_{0}t-2\alpha_{2})}\label{d2}
\end{equation}
Eq. (\ref{d1}) is integrable assuming
\begin{equation}
dq=\dfrac{1}{2\eta_{0}}\dfrac{dB}{B}
\end{equation}
with
\begin{equation}
B=\cosh(C_{1}\eta_{0}t+2\alpha_{1})+\cos(C_{2}\eta_{0}t+2\alpha_{2}).
\end{equation}
Thus, the distance between the two solitons
\begin{equation}
q(t)-q_{0}(t=0)=\dfrac{1}{2\eta_{0}}\ln\left(\dfrac{B}{B_{0}}\right)=\dfrac{1}{2\eta_{0}}\ln\left[\dfrac{\cosh(C_{1}\eta_{0}t-2\alpha_{1})+\cos(C_{2}\eta_{0}t-2\alpha_{2})}{\cosh(2\alpha_{1})+\cos(2\alpha_{2})}\right]
\end{equation}

\subsection{Adiabatic evolution of orthodromic-antidromic impulses}

In real membranes, the motion of muscles attached to the nerves can induce nerve compression, stretching, and friction, all of which intricately influence the nerve impulse profile. Moreover, nerve membranes exhibit non-uniformity in composition, with variations expected in lipid and protein content, consequently leading to localized variations in the elastic constants of nerves. As we conceptualize the membrane akin to a viscous elastic fluid, the inclusion of individual particle dynamics becomes imperative. Interactions between systems typically introduce phenomena like irreversibility and dissipation. In the previous analysis we need to incorporate the gain or loss term $\gamma (t)=\gamma_{0}$, when the impulse amplitudes become explicitly time-dependent, namely, $\eta(t)=\eta_{0}exp(2 \gamma_{0}t)$. Substituting the expression for impulses amplitudes in Eq. (\ref{c1}) yields to  

\begin{equation}
\dfrac{dY}{dt} \approx \dfrac{1}{2}\eta (t)\left[Y^{2}-\dfrac{2}{\eta_{0}}K^{2}\right], \label{geo1}
\end{equation}
being $K$ a complex constant. After integrating Eq. (\ref{geo1}) we obtain
\begin{equation}
Y=-K\sqrt{\dfrac{2}{\eta_{0}}}\tanh \left[\dfrac{\eta_{0}K}{2\gamma_{0}\sqrt{2\eta_{0}}}[\exp(2\gamma_{0}t)-1]-\alpha\right],\label{bab1}
\end{equation}
being $\alpha$ a complex constant. In a similar fashion to the case of the complex constant $C$ in Eq. (\ref{george1}) we define $K$ as

\begin{equation}
K^{2}=\dfrac{\eta_{0}}{2}\left[Y^{2}-16\eta^ {2}\exp(-2 \eta q+i\phi)-4i\gamma_{0}\dfrac{\Delta\eta}{\eta(t)}\right].\label{bab2}
\end{equation}

By a simple substitution of Eq. (\ref{bab1}) in Eq. (\ref{bab2}) we obtain the differences of soliton velocities and amplitudes

\begin{equation}
\Delta \delta=\sqrt{\dfrac{2}{\eta_{0}}}\dfrac{-K_{1}\sinh 2\psi_{1}+K_{2}\sin 2\psi_{2}}{\cosh 2\psi_{1}+\cos 2\psi_{2}},
\end{equation}
\begin{equation}
\Delta \eta=-\sqrt{\dfrac{2}{\eta_{0}}}\dfrac{K_{1}\sin 2\psi_{1}+K_{2}\sinh 2\psi_{2}}{\cosh 2\psi_{1}+\cos 2\psi_{2}},
\end{equation}
with 
\begin{equation}
\psi_{i}=\dfrac{K_{i}}{2\gamma}\sqrt{\dfrac{\eta_{0}}{2}}\left[\exp(2\gamma_{0}t)-1\right]-\alpha_{i}=\dfrac{K_{i}}{2\gamma\sqrt{2\eta_{0}}}\left[\eta (t)-\eta_{0}\right]-\alpha_{i} \quad i=1, 2, 
\end{equation}
and $K_{1}$ and $K_{2}$ being the real and complex part of $K$, respectively. The expression for the distance between solitons acquire is given by 
\begin{equation}
q(t)=q_{0}+\dfrac{1}{2\eta (t)}\ln\left[\dfrac{\cosh 2\psi_{1}+\cos 2\psi_{2}}{\cosh 2\alpha_{1}+\cos 2\alpha_{2}}\right]
\end{equation} 
The four constant parameters, $K_{i}$ and $\alpha_{i}$ with $i=1, 2$ are given by 

\begin{subequations}
\begin{equation}
\Delta \delta_{0}+i \Delta\eta_{0}=(K_{1}+iK_{2})\sqrt{\dfrac{2}{\eta_{0}}}\tanh(\alpha_{1}+i\alpha_{2})
\end{equation}
\begin{equation}
K=\pm{i}\sqrt{8\eta_{0}^ {3}\exp(-2\eta_{0}q_{0}+i\phi_{0})+2i\gamma_{0}\Delta \eta_{0}}\cosh\alpha
\end{equation}
\end{subequations}

\section{Symmetric and antisymmetric case for collision}

\subsection{Symmetric case}
\begin{figure}[H]
\begin{subfigure}{.5\textwidth}
  \centering
  \includegraphics[width=.8\linewidth]{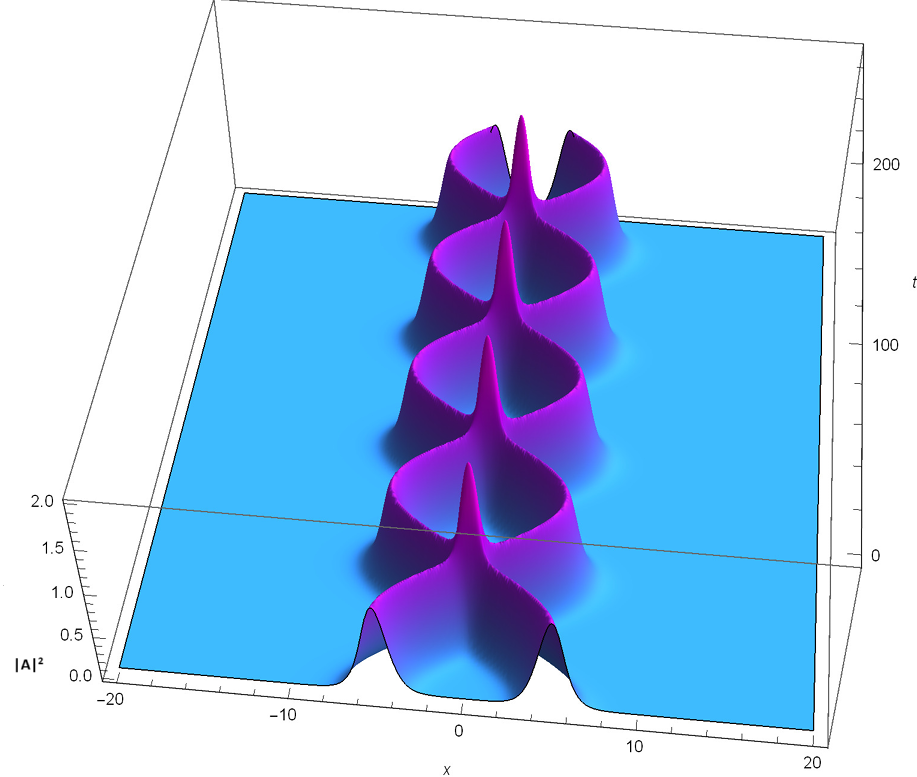}  
  \caption{Initial position $q_{0}=5$.}
  \label{fig:sub-first}
\end{subfigure}
\begin{subfigure}{.5\textwidth}
  \centering
  \includegraphics[width=.8\linewidth]{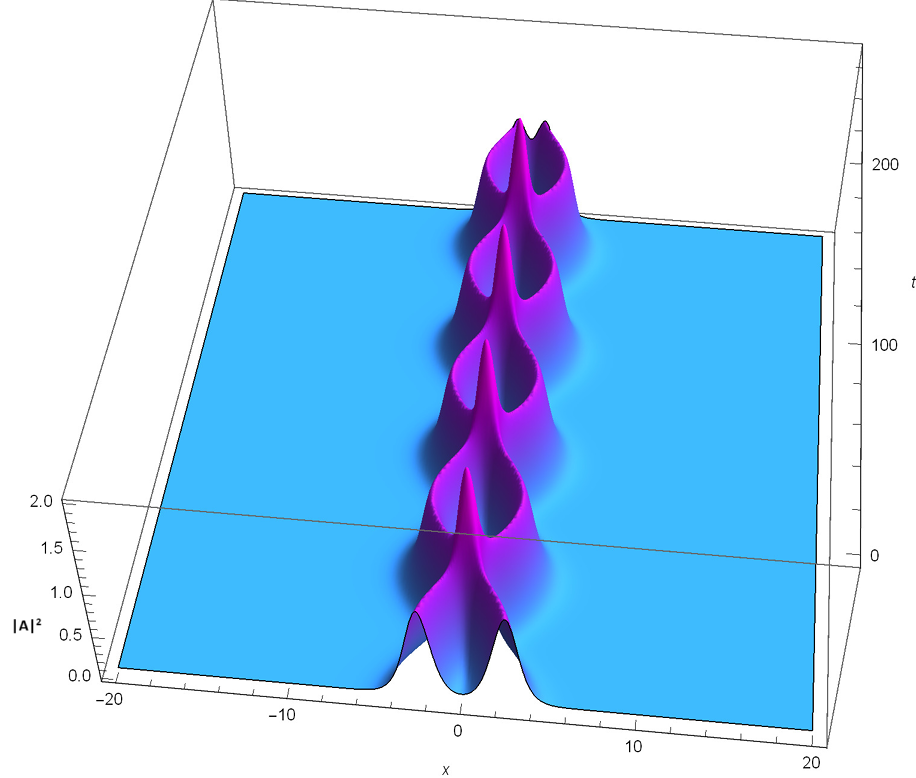}  
  \caption{Initial position $q_{0}=2.5$.}
  \label{fig:sub-second}
\end{subfigure}
\caption{{Adiabatic evolution of in phase orthodromic and antidromic impulses with $\eta_{0}=1$, $\delta_{0}=0.03$ and $\gamma_{0}=0.05$}.}
\label{fig:fig}
\end{figure}

In order to compare our analytic results with the numerical results obtained using the a variational approach we consider the case in which the wave functions are initially symmetric $(\delta \phi)$ and antisymmetric $(\Delta \phi=\pi)$.  That is, 

\begin{equation}
K_{1}=0, \quad K_{2}=\pm \sqrt{2\eta_{0}}\eta_{0}\exp(-\eta_{0}q_{0}),
\end{equation}
thus, the relative impulse separation $q(t)$
\begin{equation}
q(t)=q_{0}+\dfrac{1}{2\eta (t)}\ln \left[\cos^{2}\left(\dfrac{K_{2}(\eta(t)-\eta_{0})}{2\gamma_{0}\sqrt{2\eta_{0}}}\right)\right],\label{rodrii1}
\end{equation}
additionally to the equal initial phases of solitons, we are considering that both impulses have the same amplitude and velocity. From Eq. (\ref{rodrii1}) is easy to see that when $\gamma_{0} \to 0$, i.e., when there is no presence of viscous medium the impulse separation $q(t)$ is a periodic function and the oscillation period of the impulses is
\begin{equation}
T=\dfrac{\pi}{2\eta_{0}^ {2}}\exp(\eta_{0}q_{0}),
\end{equation} 
moreover,  it is clear to see from Eq. (\ref{rodrii1}) that the absorption at $\gamma_{0}<0$ increase the period of the impulse pair oscillation with time, which means that two impulses repeal each other, and the relative impulse separation $q(t)$ increase on the contrary case when $\gamma_{0}>0$ the period of the impulse pair oscillation with time, decrease, which means that two impulses attract each other and relative impulse separation $q(t)$ decrease as can be seen in Figure \ref{fig:fig}. 

\subsection{Asymmetric case}

For impulses with initially the same amplitudes, velocities, and opposite phases ($\phi_{0}=\pi$) we have that

\begin{equation}
K_{1}=\mp 2\sqrt{2\eta_{0}}\eta_{0}\exp(-\eta_{0}q_{0}), \quad K_{2}=0,
\end{equation}
and the relative separation $q(t)$ between the impulses becomes
\begin{equation}
q(t)=q_{0}+\dfrac{1}{2\eta (t)}\ln \left[\cosh^{2}\left(\dfrac{K_{1}(\eta(t)-\eta_{0})}{2\gamma_{0}\sqrt{2\eta_{0}}}\right)\right],\label{rodrii2}
\end{equation}

\begin{figure}[H]
\begin{subfigure}{.5\textwidth}
  \centering
  \includegraphics[width=.8\linewidth]{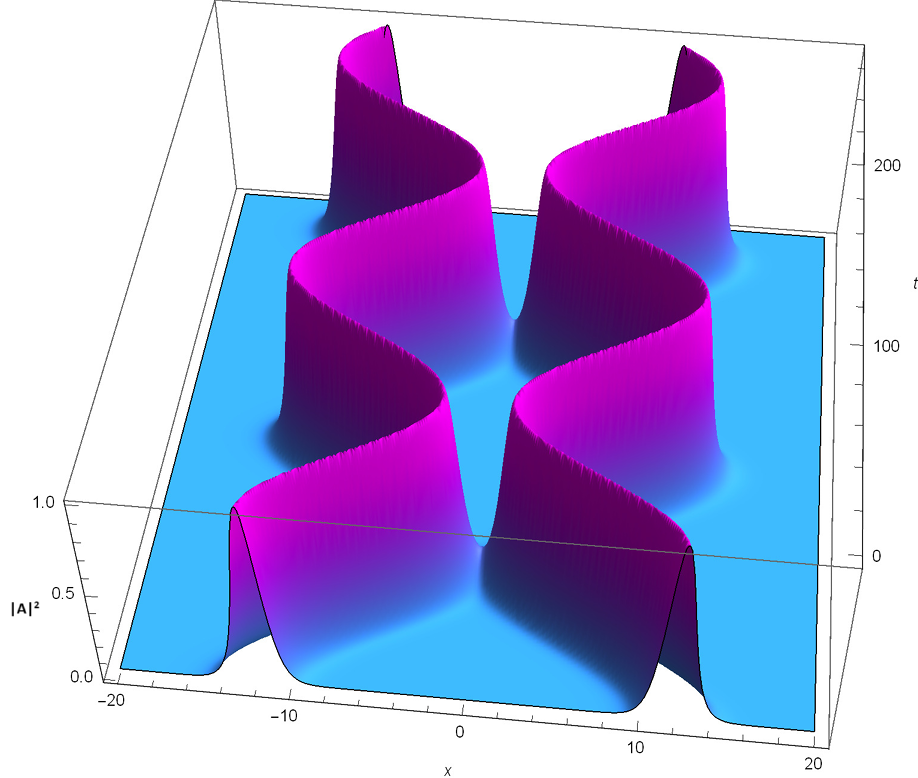}  
  \caption{Initial position $q_{0}=5$.}
  \label{fig:sub-first1}
\end{subfigure}
\begin{subfigure}{.5\textwidth}
  \centering
  \includegraphics[width=.8\linewidth]{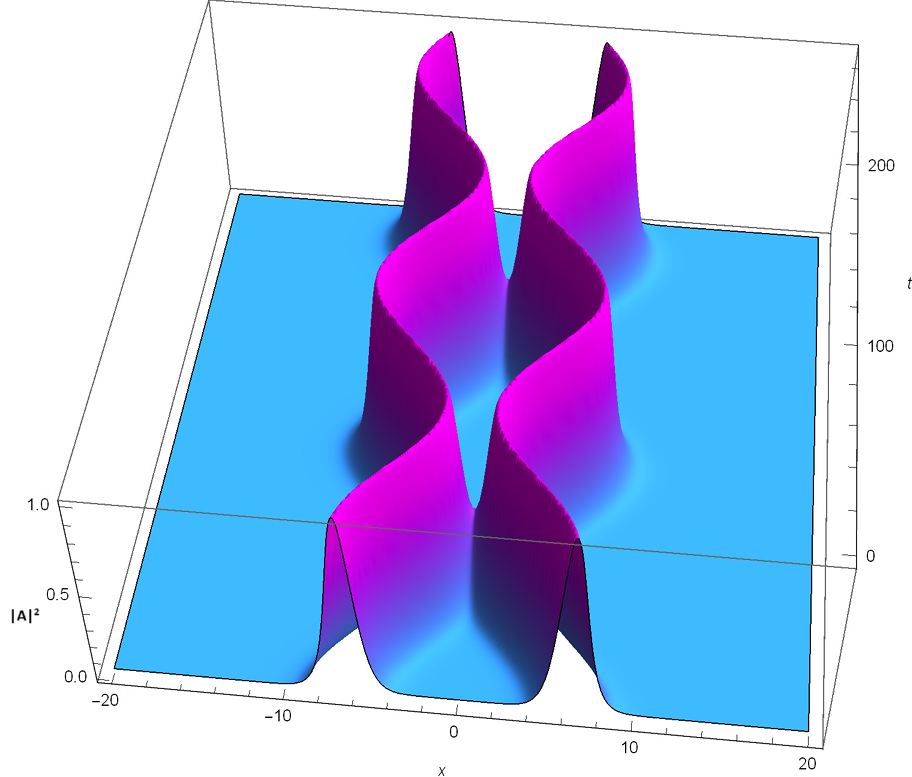}  
  \caption{Initial position $q_{0}=2.5$.}
  \label{fig:sub-second1}
\end{subfigure}
\caption{{Adiabatic evolution of out of phase orthodromic and antidromic impulses with $\eta_{0}=1$, $\delta_{0}=0.03$ and $\gamma_{0}=0.05$}.}
\label{fig1:fig1}
\end{figure}

Thus, Eq. (\ref{rodrii2}) shows that two impulses repel each other. In the presence of absorption at $\gamma_{0}<0$ the impulse separation takes place rapidly and in the case of gain at $\gamma_{0}>0$ the impulse separation takes place slowly as can be seen in Figure \ref{fig1:fig1}.  

\section{Conclusions}
In the present work we studied the interaction and adiabatic evolution of orthodromic and antidromic impulses in a viscous elastic fluid such as axoplasmic fluid. These impulses can be modelled as bright-shape solitons and their behaviour can be analysed by means of the Karpman-Maslov-Solov'ev perturbation theory. Such theory casts light on the behaviour and impact of nerve swelling or stretching when a phase-transition happens in the impulses. Starting from the well known Heimburg-Jackson model we analysed the low amplitude nonlinear excitations in a weakly dissipative thermodynamic soliton model. Using a multi-scale method to reduce the dissipative equation directly obtained from the classical model to a damped nonlinear Schrödinger equation. Consequently, we analysed the behaviour and interaction conditions to collision of two solitons directed in opposite direction. The Karpman-Maslov-Solov'ev allows to obtain information about soliton parameters once they are interacting also to understand the effect of external factors in nerve impulses. This study could be the cornerstone of future researches in nerve impulses interactions. Moreover, the damped nonlinear Schrödinger equation as a whole sub-theory within the nonlinear physics could grant new and unknown information about the electrical signal transmission and propagation in nerve fibres, which encouraged us at first to undertake the present study and further more. Every new information that can be subtracted from the thermodynamic soliton model is valuable due to the difficulties that arise from the study of such a complex neuronal system.

\section*{Acknowledgements}  This work was supported in part by funds of SIEA-UAEMEX 7036/2024 CIB Research Project. OPT acknowledges CONAHCyT by a postdoctoral fellowship.







\end{document}